# Monte Carlo-based estimation of patient absorbed dose in $^{99m}$Tc-DMSA, -MAG3, and -DTPA SPECT imaging using the University of Florida (UF) phantoms

Zeynab Khoshyari-morad[1], Reza Jahangir[2], Hashem Miri-Hakimabad[1], Najmeh Mohammadi[1], and Hossein Arabi[3]

[1] Department of Physics, Faculty of Science, Ferdowsi University of Mashhad, Mashhad, Iran.

[2] Department of Medical Radiation Engineering, Shahid Beheshti University, Tehran, Iran.

[3] Division of Nuclear Medicine and Molecular Imaging, Department of Medical Imaging, Geneva University Hospital, CH-1211 Geneva 4, Switzerland

**Abstract**

*Objective:* SPECT imaging is one of the common modalities used to examine kidney disease. The radiopharmaceuticals such as $^{99m}$Tc-MAG3, $^{99m}$Tc-DTPA, and $^{99m}$Tc-DMSA are commonly employed for pediatric patients. It is generally accepted that children's organs are more sensitive to radiation than adults, due to their growth rate. Therefore, evaluating the absorbed dose in children to avoid irrecoverable damage is highly crucial. In this work, absorbed dose by different organs of children within the SPECT imaging for the above-mentioned radiopharmaceuticals were estimated through the use of Monte Carlo simulation for patients at the ages of 4, 8, 11, and 14-years old.

*Methods:* The patient's absorbed dose was estimated based on the specific radiopharmaceutical uptake patterns and the age-related human body models. To this end, we used the University of Florida's (UF) voxel-wise baby phantoms and the Monte Carlo simulation for the pediatric patients at 4, 8, 11, and 14 years of age. The accuracy of the dosimetric outcomes was evaluated versus the ICRP 128 data.

*Results:* The highest and lowest doses absorbed by the kidneys were 0.48 mGy/MBq and 0.0042 mGy/MBq obtained for 4-year-old and 14-year-old children using $^{99m}$Tc-DMSA and $^{99m}$Tc-MAG3 radiopharmaceuticals, respectively. Moreover, in these cases, the relative errors of simulation results compared to the reference ICRP data were 12.1% and 0.95%, respectively. The simulation results were in good agreement with the ICRP 128 data.

*Conclusion:* The results showed that the highest absorbed dose was by kidneys and when $^{99m}$Tc-DMSA was used. The highest and lowest absorbed dose in the organs occurs when $^{99m}$Tc-DMSA and $^{99m}$Tc-MAG3 (in normal renal function) are used, respectively. On the other hand, the results indicated that the dose received by the organs decreases with age.

# I. Introduction

Nuclear medicine plays a key role in diagnosing a variety of diseases [1]. SPECT imaging is an essential diagnostic tool in nuclear medicine [2] which provides useful information about the body's basic biological processes [3]. SPECT images reflect the concentration of a radionuclide injected into a patient within different biological tissues [2, 4, 5]. SPECT imaging is commonly used to study kidneys and renal diseases [1, 6-9].

$^{99m}$Tc-DMSA renal scintigraphy and $^{99m}$Tc-MAG3/$^{99m}$Tc-DTPA dynamic renal scintigraphy are generally used to diagnose and monitor the response to treatment of kidney diseases [10, 11]. Determining renal uptake by radionuclide studies is a simple and non-invasive method for assessing renal function [12]. The quantitative and qualitative information obtained from renal scintigraphy depends on the radiopharmaceuticals' biological properties, mathematical models for data analysis, and experimental correlation studies [5, 13]. In addition to functional imaging, in case there is a need for structural/anatomical information, radiography and ultrasound are the methods of choice. However, if functional information is needed, nuclear medicine is preferred. In fact, there are different types of radiopharmaceuticals commonly used in clinical practice for kidney studies [11].

The biological effects of radioactive material on the body can be determined by the patient's absorbed dose [14]. In this regard, measurement of the absorbed dose in different body organs is essential for monitoring/regulating radiation hazards for patients in SPECT functional imaging [15].

Although children make up about 10% of patients undergoing radiological examinations, a small percentage of whom receive radiation therapy [15], they are highly sensitive to radiation exposure and prone to irreversible damages [16]. Thus, precise evaluation of the absorbed dose in children is crucial to avoid the side effects of SPECT imaging [17, 18].

In nuclear medicine, two main methods, containing the same theoretical considerations, are used to perform the dosimetric measurements which are developed by the International Commission on Radiological Protection (ICRP) and Medical Internal Radiation Dose (MIRD) Committee of the United States, Society of Nuclear Medicine. In the MIRD method, the absorbed dose is calculated via summing up the contributions from all main organs, while the ICRP method gives a conjecture of the radiation damage [19]. In fact, it offers recommendations for limiting radiation exposure. Dose calculations are similar in both methods, and only some of the dose indices are different [20].

Among many radioisotopes used in nuclear medicine [21], about 90% of them are utilized for diagnostic purposes. Technetium-99m ($^{99m}$Tc) is very important in clinical medicine since its half-life is 6 hours, and also its wash-out pace is relatively fast, which reduces the total patient absorbed dose. $^{99m}$Tc is suitable for labeling various pharmaceuticals, resulting in a wide range of radiopharmaceuticals [22, 23].

$^{99m}$Tc is first converted to $^{99}$Tc via gamma (141 keV) and electron emissions, then to the stable nucleus of $^{99m}$Ru. Table 1 shows the decay chain of this radioisotope and its associated probability. In this table Y(i) and E(i) are decay probability and energy, respectively.

Table 1. $^{99m}$Tc decay chain and the probability of emitted particles.

| Radiation | Y(i) (Bq.s)$^{-1}$ | E(i) (MeV) | Y(i) ×E(i) |
|---|---|---|---|
| Ce-M, γ 1 | 7.79×10$^{-01}$ | 1.629×10$^{-03}$ | 1.27×10$^{-03}$ |
| γ 2 | 8.90×10$^{-01}$ | 1.405×10$^{-01}$ | 1.25×10$^{-01}$ |
| Ce-K, γ 2 | 8.79×10$^{-02}$ | 1.195×10$^{-01}$ | 1.05×10$^{-02}$ |
| Ce-L, γ 2 | 1.07×10$^{-02}$ | 1.375×10$^{-01}$ | 1.47×10$^{-03}$ |
| Ce-M, γ 2 | 1.94×10$^{-03}$ | 1.400×10$^{-01}$ | 2.72×10$^{-04}$ |
| Ce-K, γ 3 | 6.50×10$^{-03}$ | 1.216×10$^{-01}$ | 1.94×10$^{-04}$ |
| Ce-L, γ 3 | 2.02×10$^{-03}$ | 1.396×10$^{-01}$ | 2.82×10$^{-04}$ |
| Ce-M, γ 3 | 3.96×10$^{-04}$ | 1.421×10$^{-01}$ | 5.62×10$^{-05}$ |
| Kα1 X-ray | 4.03×10$^{-02}$ | 1.837×10$^{-02}$ | 7.40×10$^{-04}$ |
| Kα2 X-ray | 2.12×10$^{-02}$ | 1.825×10$^{-02}$ | 3.87×10$^{-04}$ |
| Kβ X-ray | 1.24×10$^{-02}$ | 2.060×10$^{-02}$ | 2.55×10$^{-04}$ |
| Auger-K | 2.05×10$^{-02}$ | 1.550×10$^{-02}$ | 3.18×10$^{-04}$ |
| Auger-L | 1.04×10$^{-01}$ | 2.170×10$^{-03}$ | 2.26×10$^{-04}$ |

$^{99m}$Tc mercaptoacetyl triglycine ($^{99m}$Tc-MAG3), $^{99m}$Tc dimercaptosuccinic acid ($^{99m}$Tc-DMSA) and $^{99m}$Tc diethylene triaminepentaacetic acid ($^{99m}$Tc-DTPA) are commonly-used radiopharmaceuticals for pediatric patients with renal diseases. DMSA is a static renal scintigraphy technique, while DTPA and MAG3 are dynamic ones [11, 18, 24]. All three of these radiopharmaceuticals are suitable for renal function evaluation, but there are some differences between them. These differences are caused by the different biological properties of radiopharmaceuticals, such as renal cell retention of radioactive material, the mechanism of renal excretion, etc [7]. These radiopharmaceuticals also have a different uptake pattern within the body [15].

Over the years, there have been many attempts to perform dosimetry using the Monte Carlo code [14, 25, 26]. ICRP has also decided to use voxel phantoms to update organ dose conversion coefficients (DCCs) [27]. For Monte Carlo simulations, two sets of computational phantoms are used for defining internal organs: 1) tomographic voxel phantoms based on three-dimensional (3D) medical images, 2) stylized mathematical phantoms based on 3D surface equations. The tomographic phantoms utilize magnetic resonance (MR) or computed tomography (CT) to define human anatomy in realistic models [15].

Dosimetric calculations have been performed in different ways with mathematical models by different groups. Arteaga et al. performed dosimetric calculations for $^{99m}$Tc-DMSA, $^{99m}$Tc-DTPA, and $^{99m}$Tc-MAG3 radiopharmaceuticals using the MIRD method for newborns, 1-, 5- and 10-year-old children, as well as adults [19]. Moreover, Ebrahimnejad et al. calculated the amount of absorbed dose in the different organs for $^{99m}$Tc-DTPA Radiopharmaceutical by MIRDOSE software for 14 pediatric patients (nine female and five male) aged from 3 to 12 years [15]. Mathematical models are highly adaptable to new experimental settings; however, their anatomical descriptions of organ positions and shapes are not very realistic [28]. Voxel phantoms are the most detailed models of the human anatomy with realistic definitions and delineation of the major body organs [29].

Since the voxel phantoms are more accurate than the mathematical phantoms, and considering the recommendations by ICRP to use tomographic models instead of mathematical models [30]; in this study, the dosimetric calculations were carried out based on the voxel phantoms. In this work, the University of Florida (UF) voxel phantom has been employed for dosimetry of $^{99m}$Tc-DMSA, $^{99m}$Tc-DTPA, and $^{99m}$Tc-MAG3 radiopharmaceuticals in SPECT imaging of the pediatric patients at 4, 8, 11, and 14 years of age. The aim of the study was to determine the accurate absorbed dose in pediatric patients within the main organs. To this end, the UF phantoms, representing the pediatric patients of different ages, were employed to conduct the Monte Carlo simulation.

## II. Materials and methods

### II.1 The absorbed dose

The patients' absorbed dose is estimated through radiopharmaceutical-specific uptake patterns and accurate human body models [19]. To this end, a biokinetic model has been developed for each radiopharmaceutical that considers the radiotracer's specific metabolism and distribution within the body. Table 2 presents the biokinetic models for $^{99m}$Tc-DMSA, $^{99m}$Tc-DTPA, and $^{99m}$Tc-MAG3 radiopharmaceuticals [31], and also the organs with the highest uptakes which are regarded as the sources of radiation.

**Table 2.** Biokinetic data for $^{99m}$Tc-DMSA, $^{99m}$Tc-DTPA, and $^{99m}$Tc-MAG3.

| Organ (S) | $\tilde{A}_s/A0$ (h) | | |
|---|---|---|---|
| | 99mTc-DMSA | 99mTc-DTPA | 99mTc-MAG3 |
| Total body (excluding urinary bladder contents) | 6.8 | - | - |
| Kidneys (cortex) | 3.7 | - | - |
| Liver | 0.42 | - | - |
| Spleen | 0.042 | - | - |
| Urinary bladder contents | 0.40 | - | - |
| Normal renal function | | | |
| Total body (excluding urinary bladder contents) | - | 2.0 | - |
| Total body (excluding urinary bladder contents and kidneys) | - | - | 0.23 |
| Kidneys | - | 0.073 | 0.065 |
| Urinary bladder contents | | | |
| Adult, 15 years | - | 1.5 | 2.7 |
| 10 years | - | 1.5 | 2.3 |
| 5 years | - | 1.3 | 1.6 |
| 1 year | - | 0.83 | 1.6 |
| Abnormal renal function | | | |
| Total body (excluding urinary bladder contents) | - | 6.4 | - |
| Total body (excluding urinary bladder contents and kidneys) | - | - | 1.4 |
| Kidneys | - | 0.11 | 0.28 |
| Liver | - | - | 0.055 |
| Urinary bladder contents | | | |
| Adult, 15 years | - | 0.44 | 2.0 |
| 10 years | - | 0.44 | 1.7 |
| 5 years | - | 0.37 | 1.1 |
| 1 year | - | 0.25 | 1.1 |

The dose absorbed in the target organ is obtained by the summation of received doses from the entire source organs [19]. The mean absorbed dose $D(r_T, T_D)$ is given by:

$$D(r_T, T_D) = \sum_{r_s} \tilde{A}(r_s, T_D) S(r_T \leftarrow r_s) \qquad (1)$$

Where $T_D$ is considered 50 for adults and 70 for infants, children, and adolescents. $\tilde{A}(r_s, T_D)$ and $S(r_T \leftarrow r_s)$ are the time-integrated activity in the source organ $r_s$ such that $\tilde{A}(r_s, T_D) = \int_0^{T_D} A(r_s, t) dt$, and the mean absorbed dose rate reached the target organ $r_T$ per unit activity in source organ $r_s$ at time t after administration of the radiotracers, respectively. $S(r_T \leftarrow r_s)$ depends on the type of radiation, the emission energy, the mass of the target organ, the anatomical attributes, and the radioisotope [20].

## II.2. Monte Carlo simulation

The organs that had the highest uptakes in the radiopharmaceutical-specific biokinetic models were defined as sources for calculating the S-value in equation 1. For instance, in the DMSA radiopharmaceutical model, the renal cortex uptake was relatively high, so the right and left renal cortices were regarded as sources. Furthermore, in the DTPA radiopharmaceutical model, in which the whole kidney is considered as a source, in addition to the renal cortex, the right and left renal pelvis together with the medullar are included in the source. It should also be noted that each of these parts has different uptake values, therefore different sources with different activity rates should be considered. To this end, it is necessary to include the volume of each of these parts in the simulation. For the Monte Carlo simulation, the (e, p) mode was considered, which transfers both electrons and photons emitted from the source, rather than taking into account either one of them. The electrons and photons are subjected to various interactions within the body, such as bremsstrahlung; wherein secondary electrons and photons are generated with different energies, which should be considered within the absorbed dose calculation. Moreover, according to the type of particle emitted from the source, the energy of the radiation may vary as summarized in table 1. The absorption dose within different organs was calculated using tally + F6, which estimates the energy deposition per mass unit in MeV/g.

For accurate modeling of children's bodies, the UF baby voxel phantoms were employed which provide models for children at 4, 8, 11, and 14 years of age [17]. The 4- and 8-year-old phantoms have 74 organs/tissue, and the 11- and 14-year-old phantoms have 73 organs/tissue. Figure 1 illustrates the computational voxel phantoms for 4-, 8-,11-, and 14yearold patients. The number of source-particles that are tracked (NPS) in this simulation is $10^9$ using the Monte Carlo MCNPX code [32].

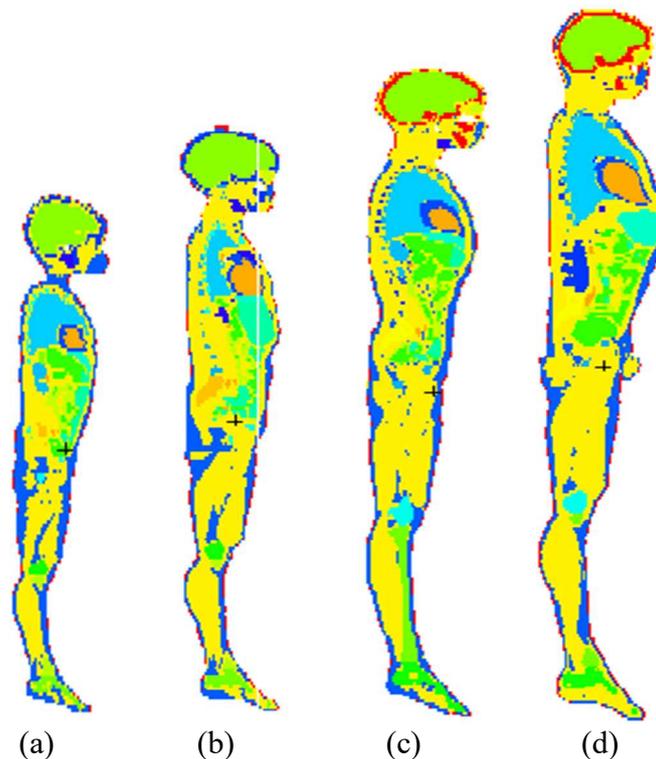

(a)   (b)   (c)   (d)

**Figure 1**. The computational phantom of (a): 4-year-old, (b): 8-year-old, (c): 11-year-old, and (d): 14-year-old childeren.

## II.3. Evaluation

Using the UF voxel phantoms in the Monte Carlo simulation, the S-value was calculated for different source organs, which include the total body (excluding urinary bladder contents), kidney (cortex), liver, spleen, urinary bladder, and kidney contents. These source organs were selected according to biokinetic models in table 2 for the different radiopharmaceuticals. For each source organ, the other organs in addition to the source organ itself are considered as target organs. Then S-value is obtained for the source organ and each of the target organs. According to the MIRD method (Eq. 1), the absorbed dose from the S-values is calculated by multiplying the S-value of each organ by its activity concentration and then summing up the results of all source organs, resulting in the radiopharmaceutical absorbed dose [31]. The details of the calculated S-values for the three radiopharmaceuticals ($^{99m}$Tc-DTPA, $^{99m}$Tc-MAG3, and $^{99m}$Tc-DMSA) are presented in Supplemental Tables 1 to 7 for the different ages.

## III. Results

Normally, the S-value is calculated before the calculation of the absorbed dose. The absorbed dose in different organs is obtained through multiplying the S-values by the total activities of the source organs. Tables 3 to 7 summarize the calculated absorbed dose (in terms of mGy/MBq) to different organs and for different age models.

The only difference between these radiopharmaceuticals is that the activity concentrations of $^{99m}$Tc-DTPA and $^{99m}$Tc-MAG3 vary in the normal and abnormal renal function, but $^{99m}$Tc-DMSA has only a single uptake pattern for both. Therefore, depending on the type of normal and abnormal kidney function, different absorbed doses are reported.

**Table 3.** Absorbed dose (mGy/MBq) from the $^{99m}$Tc-DMSA radiopharmaceutical for the different organs.

| Organ | 4y | 8y | 11y | 14y |
|---|---|---|---|---|
| Adipose | 1.13E-2 | 7.04E-3 | 6.28E-3 | 4.80E-3 |
| Skin | 6.21E-3 | 3.97E-3 | 3.41E-3 | 2.58E-3 |
| Salivary glands | 1.08E-2 | 5.97E-3 | 5.53E-3 | 4.00E-3 |
| Stomach (wall) | 2.53E-2 | 2.65E-2 | 1.88E-2 | 1.46E-2 |
| Pituitary Gland | 1.25E-2 | 6.63E-3 | 6.21E-3 | 4.27E-3 |
| Tongue | 1.13E-2 | 6.56E-3 | 6.01E-3 | 4.42E-3 |
| Tonsil | 1.21E-2 | 6.94E-3 | 6.33E-3 | 4.65E-3 |
| Brain | 1.07E-2 | 6.02E-3 | 5.59E-3 | 3.92E-3 |
| colon (wall) | 2.82E-2 | 1.90E-2 | 1.58E-2 | 1.36E-2 |
| ET2 (larynx and pharynx) and Trachea and Bronchi | 1.36E-2 | 7.75E-3 | 7.30E-3 | 5.49E-3 |
| Gall Bladder (wall) | 3.24E-2 | 2.21E-2 | 2.18E-2 | 1.47E-2 |
| active marrow (red marrow) | 1.28E-2 | 8.60E-3 | 7.79E-3 | 5.32E-3 |
| Thyroid | 1.35E-2 | 6.92E-3 | 7.91E-3 | 5.57E-3 |
| Heart | 1.88E-2 | 1.21E-2 | 1.17E-2 | 8.33E-3 |
| Liver | 4.30E-2 | 2.84E-2 | 2.63E-2 | 1.75E-2 |
| Spleen | 5.05E-2 | 3.39E-2 | 3.10E-2 | 2.51E-2 |
| Bladder (wall) | 3.15E-2 | 4.07E-2 | 2.51E-2 | 1.06E-2 |
| Small intestine (wall) | 2.81E-2 | 1.92E-2 | 1.52E-2 | 9.46E-3 |
| Esophagus | 1.98E-2 | 1.27E-2 | 1.50E-2 | 8.98E-3 |
| Pancreas | 3.87E-2 | 3.21E-2 | 3.23E-2 | 2.51E-2 |
| Thymus | 1.42E-2 | 8.87E-3 | 8.77E-3 | 6.36E-3 |
| Kidneys | 4.82E-1 | 3.08E-1 | 2.64E-1 | 2.20E-1 |
| large intestine | 2.71E-2 | 1.88E-2 | 1.53E-2 | 1.29E-2 |
| Lungs | 2.57E-2 | 1.70E-2 | 1.67E-2 | 1.37E-2 |
| Eyes | 9.38E-3 | 5.13E-3 | 4.61E-3 | 3.20E-3 |
| Adrenals | 6.07E-2 | 5.44E-2 | 4.61E-2 | 3.14E-2 |
| Remaining organs | 1.59E-2 | 9.64E-3 | 8.20E-3 | 6.41E-3 |

**Table 4.** Absorbed dose (mGy/MBq) from the $^{99m}$Tc-DTPA radiopharmaceutical for the different organs.

| Organ | 4y | 8y | 11y | 14y |
|---|---|---|---|---|
| Adipose | 3.82E-3 | 2.62E-3 | 2.28E-3 | 1.53E-3 |
| Skin | 1.81E-3 | 1.13E-3 | 9.70E-4 | 7.39E-4 |
| Salivary glands | 3.05E-3 | 1.69E-3 | 1.55E-3 | 1.13E-3 |
| Stomach ( wall ) | 4.63E-3 | 5.25E-3 | 2.61E-3 | 2.12E-3 |
| Pituitary Gland | 3.66E-3 | 1.93E-3 | 1.85E-3 | 1.24E-3 |
| Tongue | 3.17E-3 | 1.86E-3 | 1.69E-3 | 1.24E-3 |
| Tonsil | 3.40E-3 | 1.97E-3 | 1.78E-3 | 1.32E-3 |
| Brain | 3.11E-3 | 1.75E-3 | 1.62E-3 | 1.14E-3 |
| colon ( wall ) | 1.06E-2 | 8.34E-3 | 5.24E-3 | 2.36E-3 |
| ET2 (larynx and pharynx) and Trachea and Bronchi | 3.61E-3 | 2.08E-3 | 1.90E-3 | 1.42E-3 |
| Gall Bladder (wall) | 4.80E-3 | 2.81E-3 | 2.54E-3 | 1.99E-3 |
| active marrow (red marrow) | 4.15E-3 | 2.80E-3 | 2.52E-3 | 1.92E-3 |
| Thyroid | 3.59E-3 | 1.89E-3 | 2.03E-3 | 1.44E-3 |
| Heart | 4.09E-3 | 2.56E-3 | 2.31E-3 | 1.75E-3 |
| Liver | 4.45E-3 | 2.76E-3 | 2.44E-3 | 1.85E-3 |
| Spleen | 4.43E-3 | 2.77E-3 | 2.45E-3 | 2.36E-3 |
| Bladder ( wall ) | 6.21E-2 | 1.04E-1 | 5.92E-2 | 1.90E-2 |
| Small intestine ( wall ) | 6.56E-3 | 3.89E-3 | 3.28E-3 | 4.20E-3 |
| Esophagus | 4.14E-3 | 2.53E-3 | 2.47E-3 | 1.79E-3 |
| Pancreas | 5.21E-3 | 3.25E-3 | 2.95E-3 | 2.41E-3 |
| Thymus | 3.66E-3 | 2.23E-3 | 2.17E-3 | 1.60E-3 |
| Kidneys | 1.38E-2 | 8.69E-3 | 7.42E-3 | 6.13E-3 |
| large intestine | 1.24E-2 | 9.37E-3 | 7.66E-3 | 3.44E-3 |
| Lungs | 6.21E-3 | 4.05E-3 | 3.67E-3 | 2.98E-3 |
| Eyes | 2.71E-3 | 1.49E-3 | 1.33E-3 | 9.23E-4 |
| Adrenals | 5.37E-3 | 3.60E-3 | 3.14E-3 | 2.36E-3 |
| Remaining organs | 4.60E-3 | 3.15E-3 | 2.60E-3 | 1.90E-3 |

**Table 5.** Absorbed dose (mGy/MBq) from the $^{99m}$Tc-DTPA radiopharmaceutical in abnormal renal function for the different organs.

| Organ | 4y | 8y | 11y | 14y |
|---|---|---|---|---|
| Adipose | 8.96E-3 | 5.76E-3 | 4.94E-3 | 3.57E-3 |
| Skin | 4.03E-3 | 2.58E-3 | 2.24E-3 | 1.71E-3 |
| Salivary glands | 9.72E-3 | 5.40E-3 | 4.95E-3 | 3.60E-3 |
| Stomach (wall) | 1.28E-2 | 1.59E-2 | 7.84E-3 | 6.06E-3 |
| Pituitary Gland | 1.17E-2 | 6.18E-3 | 5.91E-3 | 3.97E-3 |
| Tongue | 1.01E-2 | 5.92E-3 | 5.39E-3 | 3.97E-3 |
| Tonsil | 1.08E-2 | 6.29E-3 | 5.69E-3 | 4.20E-3 |
| Brain | 9.95E-3 | 5.61E-3 | 5.20E-3 | 3.65E-3 |
| colon (wall) | 1.64E-2 | 1.08E-2 | 7.81E-3 | 5.87E-3 |
| ET2 (larynx and pharynx) and Trachea and Bronchi | 1.141E-2 | 6.61E-3 | 6.06E-3 | 4.52E-3 |
| Gall Bladder (wall) | 1.27E-2 | 8.02E-3 | 7.54E-3 | 5.65E-3 |
| active marrow (red marrow) | 9.09E-3 | 5.82E-3 | 5.16E-3 | 3.81E-3 |
| Thyroid | 1.14E-2 | 6.01E-3 | 6.45E-3 | 4.59E-3 |
| Heart | 1.26E-2 | 7.94E-3 | 7.23E-3 | 5.47E-3 |
| Liver | 1.26E-2 | 8.08E-3 | 7.30E-3 | 5.37E-3 |
| Spleen | 1.24E-2 | 7.99E-3 | 7.12E-3 | 5.84E-3 |
| Bladder (wall) | 2.76E-2 | 3.71E-2 | 2.30E-2 | 9.46E-3 |
| Small intestine (wall) | 1.32E-2 | 8.64E-3 | 7.26E-3 | 5.91E-3 |
| Esophagus | 1.28E-2 | 7.89E-3 | 7.63E-3 | 5.56E-3 |
| Pancreas | 1.39E-2 | 9.11E-3 | 8.33E-3 | 6.37E-3 |
| Thymus | 1.15E-2 | 7.02E-3 | 6.88E-3 | 5.08E-3 |
| Kidneys | 2.65E-2 | 1.69E-2 | 1.47E-2 | 1.18E-2 |
| large intestine | 1.66E-2 | 1.16E-2 | 8.52E-3 | 6.22E-3 |
| Lungs | 1.94E-2 | 1.28E-2 | 1.16E-2 | 9.35E-3 |
| Eyes | 8.65E-3 | 4.75E-3 | 4.26E-3 | 2.95E-3 |
| Adrenals | 1.46E-2 | 9.72E-3 | 8.70E-3 | 6.54E-3 |
| Remaining organs | 1.03E-2 | 6.57E-3 | 5.59E-3 | 4.20E-3 |

**Table 6.** Absorbed dose (mGy/MBq) from the $^{99m}$Tc-MAG3 radiopharmaceutical in normal renal function for the different organs.

| Organ | 4y | 8y | 11y | 14y |
|---|---|---|---|---|
| Adipose | 1.69E-3 | 1.30E-3 | 1.16E-3 | 6.44E-4 |
| Skin | 8.84E-4 | 5.35E-4 | 4.46E-4 | 3.37E-4 |
| Salivary glands | 3.74E-4 | 2.03E-4 | 1.85E-4 | 9.17E-5 |
| Stomach (wall) | 1.34E-3 | 9.73E-4 | 5.21E-4 | 4.90E-4 |
| Pituitary Gland | 4.24E-4 | 2.29E-4 | 2.10E-4 | 1.03E-4 |
| Tongue | 3.94E-4 | 2.23E-4 | 2.00E-4 | 1.07E-4 |
| Tonsil | 4.19E-4 | 2.37E-4 | 2.14E-4 | 1.14E-4 |
| Brain | 3.67E-4 | 2.05E-4 | 1.89E-4 | 9.06E-5 |
| colon (wall) | 7.98E-3 | 7.06E-3 | 4.05E-3 | 8.74E-4 |
| ET2 (larynx and pharynx) and Trachea and Bronchi | 4.71E-4 | 2.58E-4 | 2.36E-4 | 1.38E-4 |
| Gall Bladder (wall) | 1.61E-3 | 7.18E-4 | 5.37E-4 | 4.70E-4 |
| active marrow (red marrow) | 2.09E-3 | 1.52E-3 | 1.41E-3 | 1.12E-3 |
| Thyroid | 4.64E-4 | 2.34E-4 | 2.53E-4 | 1.40E-4 |
| Heart | 6.77E-4 | 4.03E-4 | 3.33E-4 | 2.18E-4 |
| Liver | 1.18E-3 | 6.26E-4 | 4.93E-4 | 4.00E-4 |
| Spleen | 1.24E-3 | 6.96E-4 | 6.00E-4 | 9.12E-4 |
| Bladder (wall) | 7.26E-2 | 1.25E-1 | 7.06E-2 | 2.19E-2 |
| Small intestine (wall) | 3.76E-3 | 1.93E-3 | 1.64E-3 | 3.33E-3 |
| Esophagus | 6.69E-4 | 3.84E-4 | 4.01E-4 | 2.30E-4 |
| Pancreas | 1.72E-3 | 9.11E-4 | 8.08E-4 | 7.89E-4 |
| Thymus | 4.97E-4 | 3.00E-4 | 2.75E-4 | 1.63E-4 |
| Kidneys | 9.44E-3 | 5.83E-3 | 4.89E-3 | 4.16E-3 |
| large intestine | 1.02E-2 | 8.10E-3 | 7.02E-3 | 2.19E-3 |
| lungs | 8.91E-4 | 5.47E-4 | 4.98E-4 | 2.48E-4 |
| Eyes | 3.25E-4 | 1.76E-4 | 1.56E-4 | 6.72E-5 |
| Adrenals | 1.71E-3 | 1.19E-3 | 9.55E-4 | 6.68E-4 |
| Remaining organs | 2.24E-3 | 1.71E-3 | 1.35E-3 | 9.02E-4 |

**Table 7.** Absorbed dose (mGy/MBq) from the $^{99m}$Tc-MAG3 radiopharmaceutical in abnormal renal function for the different organs.

| Organ | 4y | 8y | 11y | 14y |
|---|---|---|---|---|
| Adipose | 2.97E-3 | 2.03E-3 | 1.78E-3 | 9.33E-4 |
| Skin | 1.49E-3 | 9.32E-4 | 7.94E-4 | 5.97E-4 |
| Salivary glands | 2.19E-3 | 1.21E-3 | 1.11E-3 | 5.45E-4 |
| Stomach (wall) | 4.14E-3 | 4.39E-3 | 2.56E-3 | 1.73E-3 |
| Pituitary Gland | 2.57E-3 | 1.39E-3 | 1.28E-3 | 6.19E-4 |
| Tongue | 2.28E-3 | 1.33E-3 | 1.21E-3 | 6.37E-4 |
| Tonsil | 2.44E-3 | 1.42E-3 | 1.29E-3 | 6.78E-4 |
| Brain | 2.20E-3 | 1.24E-3 | 1.15E-3 | 5.48E-4 |
| colon (wall) | 9.01E-3 | 7.12E-3 | 4.61E-3 | 1.89E-3 |
| ET2 (larynx and pharynx) and Trachea and Bronchi | 2.65E-3 | 1.52E-3 | 1.40E-3 | 7.98E-4 |
| Gall Bladder (wall) | 4.99E-3 | 3.16E-3 | 2.98E-3 | 1.93E-3 |
| active marrow (red marrow) | 3.33E-3 | 2.28E-3 | 2.06E-3 | 1.55E-3 |
| Thyroid | 2.63E-3 | 1.38E-3 | 1.51E-3 | 8.07E-4 |
| Heart | 3.32E-3 | 2.09E-3 | 1.94E-3 | 1.17E-3 |
| Liver | 5.98E-3 | 3.85E-3 | 3.48E-3 | 2.24E-3 |
| Spleen | 4.82E-3 | 3.07E-3 | 2.80E-3 | 2.34E-3 |
| Bladder (wall) | 5.22E-2 | 8.79E-2 | 4.98E-2 | 1.62E-2 |
| Small intestine (wall) | 5.87E-3 | 3.55E-3 | 2.94E-3 | 3.27E-3 |
| Esophagus | 3.40E-3 | 2.10E-3 | 2.22E-3 | 1.21E-3 |
| Pancreas | 5.40E-3 | 3.79E-3 | 3.59E-3 | 2.63E-3 |
| Thymus | 2.73E-3 | 1.67E-3 | 1.63E-3 | 9.35E-4 |
| kidneys | 3.85E-2 | 2.45E-2 | 2.09E-2 | 1.75E-2 |
| large intestine | 1.04E-2 | 7.88E-3 | 6.58E-3 | 2.72E-3 |
| lungs | 4.76E-3 | 3.12E-3 | 2.90E-3 | 1.28E-3 |
| Eyes | 1.92E-3 | 1.05E-3 | 9.41E-4 | 4.05E-4 |
| Adrenals | 6.69E-3 | 5.17E-3 | 4.49E-3 | 2.91E-3 |
| Remaining organs | 3.77E-3 | 2.56E-3 | 2.11E-3 | 1.30E-3 |

As presented in Table 3, when $^{99m}$Tc-DMSA is used, the absorbed dose by the kidneys is higher for 4-year-olds (4.82E-1 mGy/MBq) and is lower in the skin for 14-year-old children (2.58E-3 mGy/MBq). Regarding Tables 4 and 5, when $^{99m}$Tc-DTPA is used, the absorbed dose is higher in the bladder wall in 8-year-olds with normal (1.04E-1 mGy/MBq) and abnormal renal function (3.71E-2 mGy/MBq), while it is lower in the skin for 14-year-old children with normal (7.39E-4 mGy/MBq) and abnormal renal function (71E-3 mGy/MBq). Similarly, in Tables 6 and 7, when $^{99m}$Tc-MAG3 is used, the absorbed dose is higher in the bladder wall for 8-year-old children with normal (1.25E-1 mGy/MBq) and abnormal renal function (8.79E-2 mGy/MBq), while the lowest absorbed dose was observed in the eyes for 14-year-old with normal (6.72E-5 mGy/MBq) and abnormal renal function (4.05E-4 mGy/MBq).

To verify the obtained results, the observations in this study have been compared with ICRP 128 data [31]. Since the data in the ICRP 128 are reported for 1-, 5-, 10-, and 15-year-old models and mathematical phantoms, a comparison was made between the results of 4, 11, and 14-year-old children in this study and the ICRP 5-, 10-, and 15-year-old models (for the three radiopharmaceuticals).

**Table 8.** Comparison of The absorbed dose for $^{99m}$Tc-DMSA, between ICRP and UF. The values are in mGy/MBq unit.

| Organ | ICRP 5 years | UF 4 years | ICRP 10 years | UF 11 years | ICRP 15 years | UF 14 years |
|---|---|---|---|---|---|---|
| Adrenals | 3.5E-2 | 6.07E-2 | 2.4E-2 | 4.61E-2 | 1.6E-2 | 3.14E-2 |
| Brain | 4.0E-3 | 1.07E-2 | 2.5E-3 | 5.59E-3 | 1.5E-3 | 3.92E-3 |
| Gallbladder wall | 2.2E-2 | 3.24E-2 | 1.4E-2 | 2.18E-2 | 1.0E-2 | 1.47E-2 |
| Stomach wall | 1.4E-2 | 2.53E-2 | 1.0E-2 | 1.88E-2 | 6.3E-3 | 1.46E-2 |
| Small intestine wall | 1.4E-2 | 2.81E-2 | 1.0E-2 | 1.52E-2 | 6.4E-3 | 9.46E-3 |
| Colon wall | 1.2E-2 | 2.82E-2 | 8.2E-3 | 1.58E-2 | 5.5E-3 | 1.36E-2 |
| Heart wall | 8.6E-3 | 1.88E-2 | 5.8E-3 | 1.17E-2 | 3.8E-3 | 8.33E-3 |
| Kidneys | 4.3E-1 | 4.82E-1 | 3.0E-1 | 2.64E-1 | 2.2E-1 | 2.20E-1 |
| Liver | 2.5E-2 | 4.30E-2 | 1.8E-2 | 2.63E-2 | 1.2E-2 | 1.75E-2 |
| Lungs | 8.0E-3 | 2.57E-2 | 5.2E-3 | 1.67E-2 | 3.5E-3 | 1.37E-2 |
| Esophagus | 5.4E-3 | 1.98E-2 | 3.4E-3 | 1.50E-2 | 2.3E-3 | 8.98E-3 |
| Pancreas | 2.3E-2 | 3.87E-2 | 1.6E-2 | 3.23E-2 | 1.1E-2 | 2.51E-2 |
| Red marrow | 9.0E-3 | 1.28E-2 | 6.8E-3 | 7.79E-3 | 4.7E-3 | 5.32E-3 |
| Skin | 4.5E-3 | 6.21E-3 | 2.9E-3 | 3.41E-3 | 1.8E-3 | 2.58E-3 |
| Spleen | 3.8E-2 | 5.05E-2 | 2.6E-2 | 3.10E-2 | 1.7E-2 | 2.51E-2 |
| Thymus | 5.4E-3 | 1.42E-2 | 3.4E-3 | 8.77E-3 | 2.3E-3 | 6.36E-3 |
| Thyroid | 5.2E-3 | 1.35E-2 | 3.1E-3 | 7.91E-3 | 1.9E-3 | 5.57E-3 |
| Urinary bladder wall | 3.1E-2 | 3.15E-2 | 2.9E-2 | 2.51E-2 | 2.3E-2 | 1.06E-2 |
| Remaining organs | 7.7E-3 | 1.59E-2 | 5.2E-3 | 8.20E-3 | 3.7E-3 | 6.41E-3 |

In table 8, for 4-, 11-, and 14-year-old children, the mean error (ME) is 1.35E-2, 4.63E-3, and 4.76E-3, the mean absolute error (MAE) is 1.35E-2, 8.83E-3, and 6.07E-3, and the root mean square error (RMSE) is 1.73E-2, 1.20E-2, and 7.49E-3, respectively.

**Table 9.** Comparison of The absorbed dose for $^{99m}$Tc-DTPA, between ICRP and UF, in normal renal function. The values are in mGy/MBq unit.

| Organ | ICRP | UF | ICRP | UF | ICRP | UF |
|---|---|---|---|---|---|---|
| | 5 years | 4 years | 10 years | 11 years | 15 years | 14 years |
| Adrenals | 4.0E-3 | 5.37E-3 | 2.7E-3 | 3.14E-3 | 1.8E-3 | 2.36E-3 |
| Brain | 2.8E-3 | 3.11E-3 | 1.7E-3 | 1.62E-3 | 1.1E-3 | 1.14E-3 |
| Gallbladder wall | 5.0E-3 | 4.80E-3 | 3.8E-3 | 2.54E-3 | 2.1E-3 | 1.99E-3 |
| Stomach wall | 4.0E-3 | 4.63E-3 | 2.8E-3 | 2.61E-3 | 1.7E-3 | 2.12E-3 |
| Small intestine wall | 7.0E-3 | 6.56E-3 | 4.9E-3 | 3.28E-3 | 3.1E-3 | 4.20E-3 |
| Colon wall | 8.1E-3 | 1.06E-2 | 6.0E-3 | 5.24E-3 | 3.9E-3 | 2.36E-3 |
| Heart wall | 3.3E-3 | 4.09E-3 | 2.2E-3 | 2.31E-3 | 1.5E-3 | 1.75E-3 |
| Kidneys | 1.1E-2 | 1.38E-2 | 7.5E-3 | 7.42E-3 | 5.3E-3 | 6.13E-3 |
| Liver | 3.8E-3 | 4.45E-3 | 2.5E-3 | 2.44E-3 | 1.6E-3 | 1.85E-3 |
| Lungs | 3.0E-3 | 6.21E-3 | 2.0E-3 | 3.67E-3 | 1.3E-3 | 2.98E-3 |
| Esophagus | 3.0E-3 | 4.14E-3 | 1.9E-3 | 2.47E-3 | 1.3E-3 | 1.79E-3 |
| Pancreas | 4.3E-3 | 5.21E-3 | 2.8E-3 | 2.95E-3 | 1.8E-3 | 2.41E-3 |
| Red marrow | 3.7E-3 | 4.15E-3 | 2.7E-3 | 2.52E-3 | 1.8E-3 | 1.92E-3 |
| Skin | 2.6E-3 | 1.81E-3 | 1.7E-3 | 9.70E-4 | 1.0E-3 | 7.39E-4 |
| Spleen | 3.9E-3 | 4.43E-3 | 2.6E-3 | 2.45E-3 | 1.6E-3 | 2.36E-3 |
| Thymus | 3.0E-3 | 3.66E-3 | 1.9E-3 | 2.17E-3 | 1.3E-3 | 1.60E-3 |
| Thyroid | 3.3E-3 | 3.59E-3 | 2.1E-3 | 2.03E-3 | 1.3E-3 | 1.44E-3 |
| Urinary bladder wall | 1.5E-1 | 6.21E-2 | 1.1E-1 | 5.92E-2 | 7.8E-2 | 1.90E-2 |
| Remaining organs | 4.2E-3 | 4.60E-3 | 3.0E-3 | 2.60E-3 | 2.1E-3 | 1.90E-3 |

In table 9, for 4-, 11-, and 14-year-old children, the ME is -3.83E-3, -2.80E-3, and -2.82E-3, the MAE is 5.58E-3, 3.14E-3, and 3.61E-3, and the RMSE is 2.02E-2, 1.17E-2, and 1.35E-2, respectively.

**Table 10.** Comparison of The absorbed dose for $^{99m}$Tc-DTPA, between ICRP and UF, in abnormal renal function. The values are in mGy/MBq unit.

| Organ | ICRP | UF | ICRP | UF | ICRP | UF |
|---|---|---|---|---|---|---|
| | 5 years | 4 years | 10 years | 11 years | 15 years | 14 years |
| Adrenals | 1.1E-2 | 1.46E-2 | 7.6E-3 | 8.70E-3 | 5.1E-3 | 6.54E-3 |
| Brain | 9.1E-3 | 9.95E-3 | 5.7E-3 | 5.20E-3 | 3.5E-3 | 3.65E-3 |
| Gallbladder wall | 1.3E-2 | 1.27E-2 | 9.2E-3 | 7.54E-3 | 5.7E-3 | 5.65E-3 |
| Stomach wall | 1.1E-2 | 1.28E-2 | 7.9E-3 | 7.84E-3 | 5.0E-3 | 6.06E-3 |
| Small intestine wall | 1.3E-2 | 1.32E-2 | 8.5E-3 | 7.26E-3 | 5.6E-3 | 5.91E-3 |
| Colon wall | 1.3E-2 | 1.64E-2 | 8.7E-3 | 7.81E-3 | 5.8E-3 | 5.87E-3 |
| Heart wall | 1.0E-2 | 1.26E-2 | 7.0E-3 | 7.23E-3 | 4.7E-3 | 5.47E-3 |
| Kidneys | 1.9E-2 | 2.65E-2 | 1.3E-2 | 1.47E-2 | 9.2E-3 | 1.18E-2 |
| Liver | 1.1E-2 | 1.26E-2 | 7.1E-3 | 7.30E-3 | 4.6E-3 | 5.37E-3 |
| Lungs | 9.5E-3 | 1.94E-2 | 6.2E-3 | 1.16E-2 | 4.2E-3 | 9.35E-3 |
| Esophagus | 9.6E-3 | 1.28E-2 | 6.2E-3 | 7.63E-3 | 4.2E-3 | 5.56E-3 |
| Pancreas | 1.2E-2 | 1.39E-2 | 8.0E-3 | 8.33E-3 | 5.3E-3 | 6.37E-3 |
| Red marrow | 9.3E-3 | 9.09E-3 | 6.4E-3 | 5.16E-3 | 4.2E-3 | 3.81E-3 |
| Skin | 6.7E-3 | 4.03E-3 | 4.2E-3 | 2.24E-3 | 2.6E-3 | 1.71E-3 |
| Spleen | 1.1E-2 | 1.24E-2 | 7.3E-3 | 7.12E-3 | 4.7E-3 | 5.84E-3 |
| Thymus | 9.6E-3 | 1.15E-2 | 6.2E-3 | 6.88E-3 | 4.2E-3 | 5.08E-3 |
| Thyroid | 1.1E-2 | 1.14E-2 | 6.7E-3 | 6.45E-3 | 4.2E-3 | 4.59E-3 |
| Urinary bladder wall | 5.0E-2 | 2.76E-2 | 3.9E-2 | 2.30E-2 | 2.7E-2 | 9.46E-3 |
| Remaining organs | 9.7E-3 | 1.03E-2 | 6.3E-3 | 5.59E-3 | 4.1E-3 | 4.20E-3 |

In table 10, for 4-, 11-, and 14-year-old children, the ME is 8.04E-4, -7.17E-4, and -8.47E-5, the MAE is 3.50E-3, 1.88E-3, and 1.90E-3, and RMSE is 6.16E-3, 3.99E-3, and 4.30E-3, respectively.

**Table 11.** Comparison of The absorbed dose for $^{99m}$Tc-MAG3, between ICRP and UF. The values are in mGy/MBq unit.

| Organ | ICRP | UF | ICRP | UF | ICRP | UF |
|---|---|---|---|---|---|---|
| | 5 years | 4 years | 10 years | 11 years | 15 years | 14 years |
| Adrenals | 1.2E-3 | 1.71E-3 | 8.2E-4 | 9.55E-4 | 5.1E-4 | 6.68E-4 |
| Brain | 3.5E-4 | 3.67E-4 | 2.2E-4 | 1.89E-4 | 1.3E-4 | 9.06E-5 |
| Gallbladder wall | 1.7E-3 | 1.61E-3 | 2.0E-3 | 5.37E-4 | 8.7E-4 | 4.70E-4 |
| Stomach wall | 1.3E-3 | 1.34E-3 | 9.7E-4 | 5.21E-4 | 4.9E-4 | 4.90E-4 |
| Small intestine wall | 4.6E-3 | 3.76E-3 | 4.2E-3 | 1.64E-3 | 3.0E-3 | 3.33E-3 |
| Colon wall | 6.0E-3 | 7.98E-3 | 5.9E-3 | 4.05E-3 | 4.3E-3 | 8.74E-4 |
| Heart wall | 5.7E-4 | 6.77E-4 | 3.7E-4 | 3.33E-4 | 2.4E-4 | 2.18E-4 |
| Kidneys | 8.4E-3 | 9.44E-3 | 5.9E-3 | 4.89E-3 | 4.2E-3 | 4.16E-3 |
| Liver | 1.1E-3 | 1.18E-3 | 7.5E-4 | 4.93E-4 | 4.3E-4 | 4.00E-4 |
| Lungs | 5.0E-4 | 8.91E-4 | 3.3E-4 | 4.98E-4 | 2.1E-4 | 2.48E-4 |
| Esophagus | 4.4E-4 | 6.69E-4 | 2.8E-4 | 4.01E-4 | 1.8E-4 | 2.30E-4 |
| Pancreas | 1.3E-3 | 1.72E-3 | 9.3E-4 | 8.08E-4 | 5.0E-4 | 7.89E-4 |
| Red marrow | 1.5E-3 | 2.09E-3 | 1.6E-3 | 1.41E-3 | 1.2E-3 | 1.12E-3 |
| Skin | 9.7E-4 | 8.84E-4 | 8.3E-4 | 4.46E-4 | 5.7E-4 | 3.37E-4 |
| Spleen | 1.2E-3 | 1.24E-3 | 7.9E-4 | 6.00E-4 | 4.9E-4 | 9.12E-4 |
| Thymus | 4.4E-4 | 4.97E-4 | 2.8E-4 | 2.75E-4 | 1.8E-4 | 1.63E-4 |
| Thyroid | 4.4E-4 | 4.64E-4 | 2.7E-4 | 2.53E-4 | 1.6E-4 | 1.40E-4 |
| Urinary bladder wall | 1.8E-1 | 7.26E-2 | 1.7E-1 | 7.06E-2 | 1.4E-1 | 2.19E-2 |
| Remaining organs | 2.2E-3 | 2.24E-3 | 2.1E-3 | 1.35E-3 | 1.6E-3 | 9.02E-4 |

In table 11, for 4-, 11-, and 14-year-old children, the ME is -5.41E-3, -5.74E-3, and -6.41E-3, the MAE is 6.00E-3, 5.74E-3, and 6.55E-3, and the RMSE is 2.46E-2, 2.28E-2, and 2.71E-2, respectively.

**Table 12.** Comparison of The absorbed dose for $^{99m}$Tc-MAG3, between ICRP and UF, in abnormal renal function. The values are in mGy/MBq unit.

| Organ | ICRP | UF | ICRP | UF | ICRP | UF |
|---|---|---|---|---|---|---|
| | 5 years | 4 years | 10 years | 11 years | 15 years | 14 years |
| Adrenals | 4.8E-3 | 6.69E-3 | 3.2E-3 | 4.49E-3 | 2.1E-3 | 2.91E-3 |
| Brain | 2.0E-3 | 2.20E-3 | 1.3E-3 | 1.15E-3 | 7.7E-4 | 5.48E-4 |
| Gallbladder wall | 4.6E-3 | 4.99E-3 | 3.8E-3 | 2.98E-3 | 2.2E-3 | 1.93E-3 |
| Stomach wall | 3.5E-3 | 4.14E-3 | 2.6E-3 | 2.56E-3 | 1.5E-3 | 1.73E-3 |
| Small intestine wall | 6.0E-3 | 5.87E-3 | 5.0E-3 | 2.94E-3 | 3.5E-3 | 3.27E-3 |
| Colon wall | 6.9E-3 | 9.01E-3 | 6.1E-3 | 4.61E-3 | 4.4E-3 | 1.89E-3 |
| Heart wall | 2.7E-3 | 3.32E-3 | 1.8E-3 | 1.94E-3 | 1.2E-3 | 1.17E-3 |
| Kidneys | 3.4E-2 | 3.85E-2 | 2.4E-2 | 2.09E-2 | 1.7E-2 | 1.75E-2 |
| Liver | 3.8E-3 | 5.98E-3 | 2.7E-3 | 3.48E-3 | 1.8E-3 | 2.24E-3 |
| Lungs | 2.4E-3 | 4.76E-3 | 1.6E-3 | 2.90E-3 | 1.1E-3 | 1.28E-3 |
| Esophagus | 2.3E-3 | 3.40E-3 | 1.5E-3 | 2.22E-3 | 9.7E-4 | 1.21E-3 |
| Pancreas | 4.3E-3 | 5.40E-3 | 2.9E-3 | 3.59E-3 | 1.9E-3 | 2.63E-3 |
| Red marrow | 3.1E-3 | 3.33E-3 | 2.6E-3 | 2.06E-3 | 1.9E-3 | 1.55E-3 |
| Skin | 2.0E-3 | 1.49E-3 | 1.5E-3 | 7.94E-4 | 9.6E-4 | 5.97E-4 |
| Spleen | 4.3E-3 | 4.82E-3 | 2.9E-3 | 2.80E-3 | 1.9E-3 | 2.34E-3 |
| Thymus | 2.3E-3 | 2.73E-3 | 1.5E-3 | 1.63E-3 | 9.7E-4 | 9.35E-4 |
| Thyroid | 2.4E-3 | 2.63E-3 | 1.5E-3 | 1.51E-3 | 9.5E-4 | 8.07E-4 |
| Urinary bladder wall | 1.3E-1 | 5.22E-2 | 1.3E-1 | 4.98E-2 | 1.1E-1 | 1.62E-2 |
| Remaining organs | 3.4E-3 | 3.77E-3 | 2.8E-3 | 2.11E-3 | 2.1E-3 | 1.30E-3 |

In table 12, for 4-, 11-, and 14-year-old children the ME is -3.14E-3, -4.47E-3, and -5.01E-3, the MAE is 5.12E-3, 5.00E-3, and 5.38E-3, and the RMSE is 1.79E-2, 1.84E-2, and 2.15E-2, respectively.

For better comparison, figure 2 depicts the absorbed dose from the three radiopharmaceuticals at different ages and in different important target organs such as the kidneys, liver, brain, and lungs.

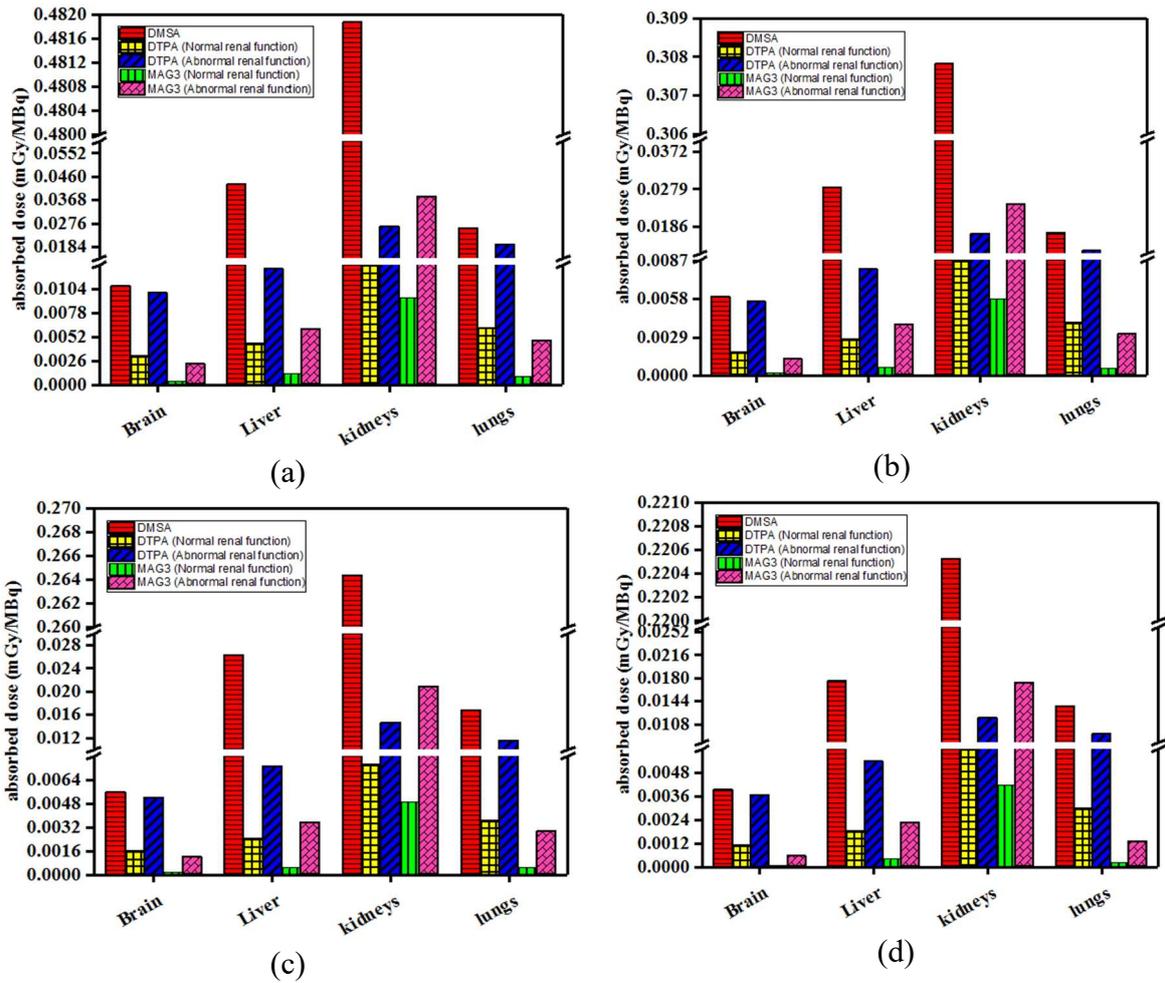

**Figure 2.** Absorbed dose from the different radiopharmaceuticals within the different target organs of (a): 4-year-old, (b): 8-year-old, (c): 11-year-old, and (d): 14-year-old models.

Moreover, to investigate the effect of age on the absorbed dose, the absorbed dose of several target organs related to different ages from the three radiopharmaceuticals is illustrated in figure 3.

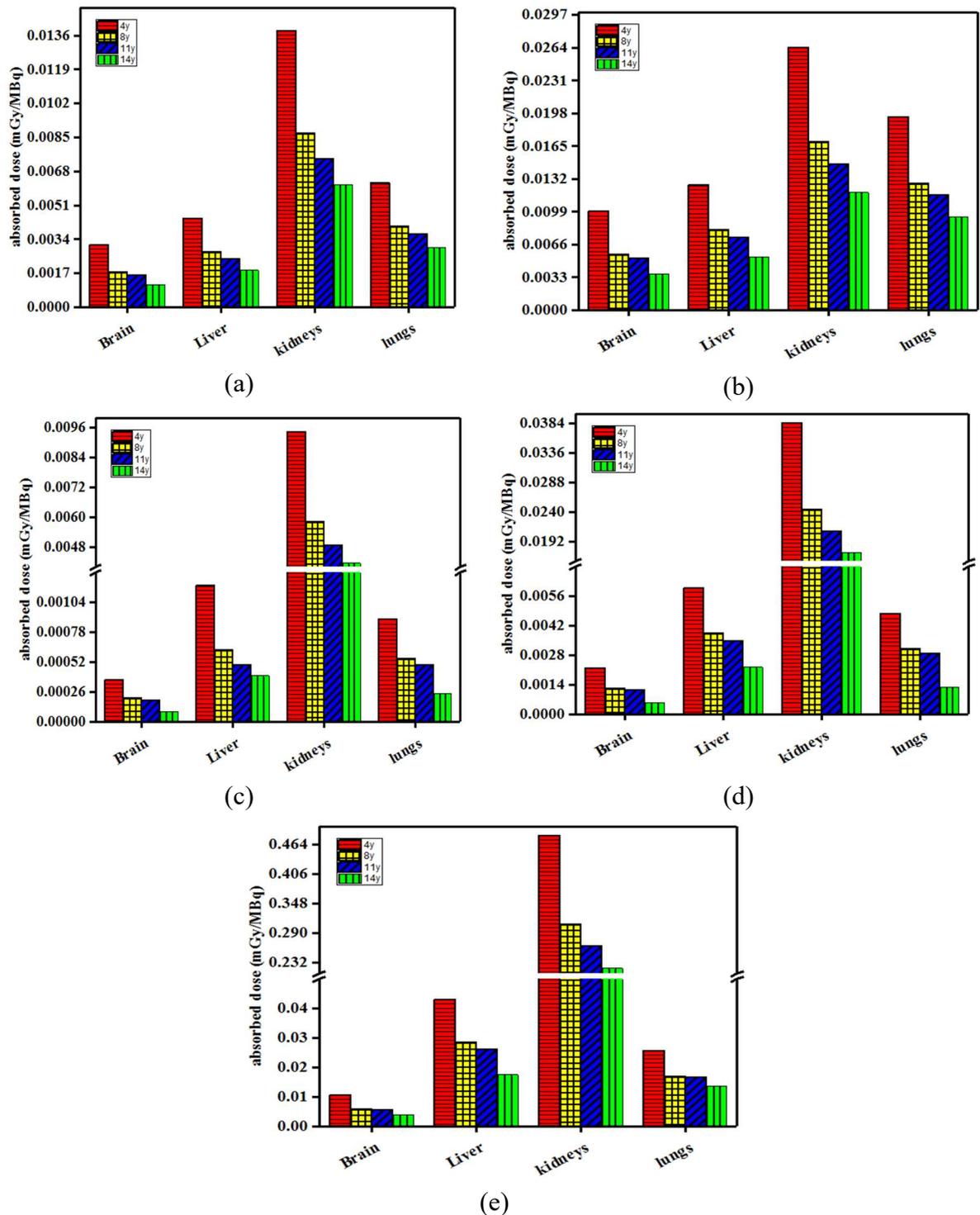

**Figure 3.** Absorbed dose from (a): $^{99m}$Tc-DTPA radiopharmaceutical, in normal renal function, (b): $^{99m}$Tc-DTPA radiopharmaceutical, in abnormal renal function, (c): $^{99m}$Tc-MAG3 radiopharmaceutical, in normal renal function, (d): $^{99m}$Tc-MAG3 radiopharmaceutical, in abnormal renal function and (e): $^{99m}$Tc-DMSA radiopharmaceuticals for the different ages and target organs

In each of the figures above, the type of radiopharmaceutical is fixed, but the age is different. It's clear that an increase in age would result in an increase in the distance between organs of the body, and consequently, reduces the absorbed dose in the organs.

The absorbed dose was calculated for different organs. Due to different energy ranges of the electrons and photons emitted by the $^{99m}$Tc radioisotope and the type of interactions, their contribution to the absorbed dose and/or energy within the different organs will be different. Figure 4 illustrates the S-values calculated for photons and electrons within the target organs of the kidneys, lungs, and brain of a 4-year-old child when the kidney is considered the source.

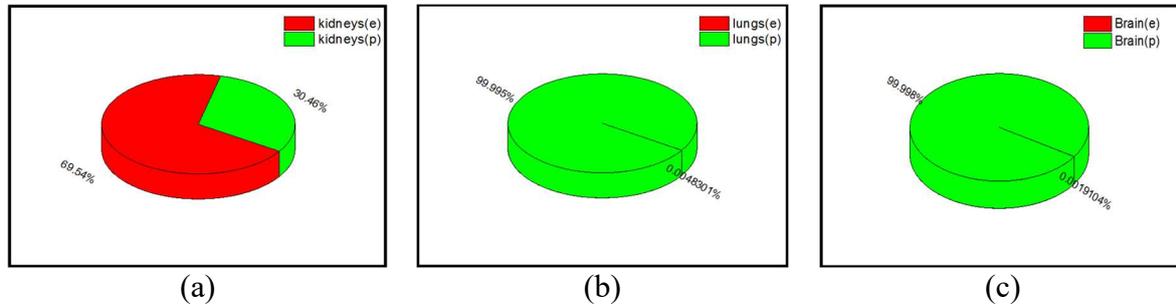

(a)    (b)    (c)

**Figure 4.** The contribution of electrons and photons in the S-value for the source organ of the kidneys and the target organ of (a): kidneys, (b): lungs, and (c): Brain.

**IV. Discission**

As figure 4 (a) presents, the contribution of electrons and photons in the S-value for the source organ of the kidneys, and the target organs of the kidneys is 69.54% and 30.46%. In figure 4 (b), the contribution of electrons and photons in the S-value for the source organs of the kidneys and the target organ of lung is 0.005% and 99.995%, respectively. In figure 4 (c), the contribution of electrons and photons in the S-value for the source organ of kidney and the target organ of brain is 0.002% and 99.998%, respectively. It is clear that as the target organ moves away from the source organ, the contribution of electrons in the absorbed dose is greatly reduced. Since the electron is a charged particle, it interacts and loses energy while traveling within the medium. Thus, the closer the target organ to the source organ, the more energy received from the electrons. For the distant organ, the contribution of the electrons is approximately zero in the corresponding S-value. Most of the electron's contribution, which is more than twice of the photon's, is related to the organ of the kidney itself because of the greater self-dose.

As observed in Table 2, for $^{99m}$Tc-DMSA radiopharmaceutical, the highest activity concentration from the radiotracer is in the kidney organ. Also, as we know, the highest amount of absorbed dose in each organ is associated with self-dose. As indicated in Table 3, the maximum absorbed dose belongs to the kidney which is in agreement with our observation in this study. The same observation was made for the radiopharmaceuticals. For $^{99m}$Tc-DTPA and $^{99m}$Tc-MAG3, the highest absorption occurred in the bladder, which is evident in Tables 4 to 7. However, considering the 14-year-old model, there is an exception in both of these radiopharmaceuticals where the absorbed dose of the kidney is slightly higher than that of the bladder.

In general, the absorbed dose decreases with age, but there are exceptions where the absorbed dose increases with age for $^{99m}$Tc-DMSA radiopharmaceutical in the stomach wall, colon wall, bladder wall (in which the absorbed dose of the 8-year-old child is greater than those of 4-year-old), thyroid, esophagusand pancreas (in which the absorbed dose of the 11-year-old child is greater than those of the 8-year-old) (Table 3). Regarding the normal renal function model with $^{99m}$Tc-DTPA (Table 4), these exceptions include the stomach wall and bladder wall (in which the absorbed dose of the 8-year-old child is greater than those of the 4-year-old), the thyroid (in which the absorbed dose of the 11-year-old child is greater than those of the 8-year-old), and the small intestinal wall (in which the absorbed dose of the 14-year-old child is greater than those of the 11-year-old). For abnormal renal function in Table 5, these exceptions include the stomach wall, bladder wall (in which the absorbed dose of the 8-year-old child is greater than those of the 4-year-old), and the thyroid (in which the absorbed dose of the 11-year-old child is greater than those of the 8-year-old). Similarly, In $^{99m}$Tc-MAG3 for normal renal function (Table 6), these exceptions are the bladder wall (in which the absorbed dose of the 8-year-old child is greater than those of the 4-year-old), the thyroid and esophagus (in which the absorbed dose of the 11-year-old child is greater than those of the 8-year-old), the spleen and the small intestinal wall (in which the absorbed dose of the 14-year-old child is greater than those of the 11-year-old). For the abnormal renal function (Table 7), these exceptions include the stomach wall and bladder wall (in which the absorbed dose of the 8-year-old child is greater than those of the 4-year-old), the thyroid and esophagus (in which the absorbed dose of the 11-year-old child is greater than those of the 8-year-old), and the small intestinal wall (in which the absorbed dose of the 14-year-old child is greater than those of the 11-year-old).

As shown in Figure 3, for the entire ages, the highest and lowest value of the organ absorbed dose (in normal renal function) occurs when $^{99m}$TC-DMSA and $^{99m}$Tc-MAG3 are used, respectively.

Moreover, when the renal function is abnormal, the results suggest that it would be better to reduce the dose of the radiotracer. This is also recommended by Mendichovszky et al. [18].

Arteaga et al. also utilized the MIRD method to calculate dosimetry of $^{99m}$Tc (DMSA, DTPA, and MAG3) for newborns and children within the source and target organs, and reported that the lowest dose to the kidneys was observed when using $^{99m}$Tc-MAG3, and the largest when using $^{99m}$Tc-DMSA [19]. The trend of the absorbed dose observed in this study for the 5- and 10-year-old, as well as the adult model is consistent with our observation.

Regarding Tables 8 to 12, the absorbed dose calculated in this study and the data from ICRP128 are in good agreement. The discrepancy between the ICRP128 data and the results observed in this study is due to the differences in the pediatric phantoms and subsequently, different organ volumes and locations. Phantom voxel has a more realistic description of the shape, volume, and location of the organs [17], and the absorbed dose strongly depends on the distances between the organs.

This article employed the MCNPX code which bears some limitations including the amount of NPS and the minimum definable/traceable energy levels (however, this limitation did not affect the accuracy of the results obtained in this study) [32]. Though the voxel phantoms provide accurate anatomical details, they are not able to model internal organ motion (such as respiratory motion), which warrants a separate thorough investigation [28, 30].

**V. Conclusion**

Using the Monte Carlo simulation framework, the MIRD method, and the UF voxel phantoms, the absorbed dose was calculated for the pediatric models at the ages of 4, 8, 11, and 14 in SPECT imaging with $^{99m}$Tc-DMSA, $^{99m}$Tc-DTPA, and $^{99m}$Tc-MAG3 radiopharmaceuticals. The results indicated that the highest and lowest organ absorbed dose occurred when $^{99m}$Tc-DMSA and $^{99m}$Tc-MAG3 radiopharmaceuticals were used in normal renal function, respectively.

The organ absorbed dose decreased in all radiopharmaceuticals as the age increased owing to the larger distances between the source and target organs. The results suggest the injected dose could be reduced when the patient has abnormal renal function.

# Supplemental materials

**Supplemental Table 1.** S-value for the organ of the source of the kidney (cortex) in units of mGy/Bq.s

| Organ | 4y | 8y | 11y | 14y |
|---|---|---|---|---|
| Adipose | 1.25E-13 | 6.40E-14 | 7.18E-14 | 6.95E-14 |
| Skin | 1.32E-13 | 8.33E-14 | 7.04E-14 | 5.30E-14 |
| Salivary glands | 3.00E-14 | 1.39E-14 | 1.73E-14 | 1.19E-14 |
| Stomach (wall) | 7.82E-13 | 6.29E-13 | 6.92E-13 | 5.59E-13 |
| Pituitary Gland | 1.43E-14 | 4.72E-15 | 6.19E-15 | 4.72E-15 |
| Tongue | 3.55E-14 | 1.61E-14 | 1.70E-14 | 1.36E-14 |
| Tonsil | 3.34E-14 | 1.58E-14 | 1.54E-14 | 1.37E-14 |
| Brain | 9.19E-15 | 3.64E-15 | 4.37E-15 | 2.68E-15 |
| colon (wall) | 7.79E-13 | 5.51E-13 | 5.34E-13 | 4.96E-13 |
| ET2 (larynx and pharynx) and Trachea and Bronchi | 9.26E-14 | 4.37E-14 | 5.51E-14 | 4.83E-14 |
| Gall Bladder (wall) | 1.09E-12 | 7.15E-13 | 7.07E-13 | 4.02E-13 |
| active marrow (red marrow) | 2.19E-13 | 1.70E-13 | 1.64E-13 | 9.16E-14 |
| Thyroid | 7.80E-14 | 3.24E-14 | 6.50E-14 | 4.85E-14 |
| Heart | 3.10E-13 | 2.18E-13 | 2.25E-13 | 1.59E-13 |
| Liver | 1.20E-12 | 7.80E-13 | 7.26E-13 | 4.22E-13 |
| Spleen | 2.00E-12 | 1.34E-12 | 1.32E-12 | 1.05E-12 |
| Bladder (wall) | 1.32E-13 | 5.34E-14 | 2.53E-14 | 3.42E-14 |
| Small intestine (wall) | 1.03E-12 | 7.39E-13 | 5.43E-13 | 2.08E-13 |
| Esophagus | 3.68E-13 | 2.60E-13 | 4.50E-13 | 2.12E-13 |
| Pancreas | 1.66E-12 | 1.60E-12 | 1.69E-12 | 1.34E-12 |
| Thymus | 1.14E-13 | 8.49E-14 | 8.96E-14 | 6.69E-14 |
| Kidneys | 3.50E-11 | 2.24E-11 | 1.92E-11 | 1.61E-11 |
| Large intestine | 6.80E-13 | 4.66E-13 | 4.40E-13 | 4.24E-13 |
| Lungs | 3.04E-13 | 1.99E-13 | 2.74E-13 | 2.59E-13 |
| Eyes | 1.21E-14 | 5.13E-15 | 5.57E-15 | 3.94E-15 |
| Adrenals | 3.27E-12 | 3.26E-12 | 2.70E-12 | 1.84E-12 |
| Remaining organs | 3.54E-13 | 1.89E-13 | 1.60E-13 | 1.38E-13 |

**Supplemental Table 2.** S-value for the organ of the source of the kidney in units of mGy/Bq.s

| Organ | 4y | 8y | 11y | 14y |
|---|---|---|---|---|
| Adipose | 1.24E-13 | 6.37E-14 | 7.15E-14 | 6.92E-14 |
| Skin | 1.31E-13 | 8.29E-14 | 7.01E-14 | 5.28E-14 |
| Salivary glands | 3.01E-14 | 1.42E-14 | 1.71E-14 | 1.17E-14 |
| Stomach (wall) | 7.84E-13 | 6.29E-13 | 6.77E-13 | 5.54E-13 |
| Pituitary Gland | 1.32E-14 | 4.79E-15 | 5.59E-15 | 6.35E-15 |
| Tongue | 3.56E-14 | 1.60E-14 | 1.66E-14 | 1.38E-14 |
| Tonsil | 3.41E-14 | 1.58E-14 | 1.48E-14 | 1.35E-14 |
| Brain | 9.25E-15 | 3.59E-15 | 4.31E-15 | 2.66E-15 |
| colon (wall) | 7.60E-13 | 5.45E-13 | 5.27E-13 | 4.92E-13 |
| ET2 (larynx and pharynx) and Trachea and Bronchi | 9.27E-14 | 4.39E-14 | 5.40E-14 | 4.78E-14 |
| Gall Bladder (wall) | 1.10E-12 | 7.19E-13 | 7.09E-13 | 4.01E-13 |
| active marrow (red marrow) | 2.17E-13 | 1.69E-13 | 1.63E-13 | 9.14E-14 |
| Thyroid | 7.52E-14 | 3.25E-14 | 6.26E-14 | 4.86E-14 |
| Heart | 3.12E-13 | 2.18E-13 | 2.21E-13 | 1.58E-13 |
| Liver | 1.21E-12 | 7.71E-13 | 7.18E-13 | 4.21E-13 |
| Spleen | 2.00E-12 | 1.34E-12 | 1.28E-12 | 1.06E-12 |
| Bladder (wall) | 1.32E-13 | 5.34E-14 | 2.53E-14 | 3.39E-14 |
| Small intestine (wall) | 1.02E-12 | 7.36E-13 | 5.40E-13 | 2.08E-13 |
| Esophagus | 3.70E-13 | 2.58E-13 | 4.36E-13 | 2.10E-13 |
| Pancreas | 1.65E-12 | 1.59E-12 | 1.65E-12 | 1.33E-12 |
| Thymus | 1.15E-13 | 8.45E-14 | 8.80E-14 | 6.73E-14 |
| Kidneys | 3.59E-11 | 2.29E-11 | 1.96E-11 | 1.64E-11 |
| Large intestine | 6.64E-13 | 4.62E-13 | 4.35E-13 | 4.21E-13 |
| Lungs | 3.06E-13 | 1.98E-13 | 2.68E-13 | 2.56E-13 |
| Eyes | 1.20E-14 | 5.06E-15 | 5.51E-15 | 3.83E-15 |
| Adrenals | 3.15E-12 | 3.04E-12 | 2.55E-12 | 1.77E-12 |
| Remaining organs | 3.46E-13 | 1.87E-13 | 1.58E-13 | 1.37E-13 |

**Supplemental Table 3**. S-value for the organ of the source of the Liver in units of mGy/Bq.s

| Organ | 4y | 8y | 11y | 14y |
|---|---|---|---|---|
| Adipose | 1.24E-13 | 5.16E-14 | 6.14E-14 | 6.11E-14 |
| Skin | 1.38E-13 | 8.91E-14 | 7.14E-14 | 4.91E-14 |
| Salivary glands | 6.01E-14 | 3.02E-14 | 3.11E-14 | 1.16E-14 |
| Stomach (wall) | 8.70E-13 | 8.42E-13 | 8.81E-13 | 5.05E-13 |
| Pituitary Gland | 2.87E-14 | 1.09E-14 | 1.42E-14 | 3.95E-15 |
| Tongue | 8.27E-14 | 4.12E-14 | 3.55E-14 | 1.59E-14 |
| Tonsil | 6.70E-14 | 3.33E-14 | 3.06E-14 | 1.26E-14 |
| Brain | 1.66E-14 | 6.84E-15 | 7.51E-15 | 2.56E-15 |
| colon (wall) | 4.49E-13 | 2.43E-13 | 3.95E-13 | 5.43E-13 |
| ET2 (larynx and pharynx) and Trachea and Bronchi | 1.92E-13 | 9.50E-14 | 9.54E-14 | 4.43E-14 |
| Gall Bladder (wall) | 3.16E-12 | 2.84E-12 | 3.11E-12 | 2.28E-12 |
| active marrow (red marrow) | 1.79E-13 | 1.29E-13 | 1.06E-13 | 4.85E-14 |
| Thyroid | 1.62E-13 | 7.29E-14 | 1.15E-13 | 4.42E-14 |
| Heart | 8.89E-13 | 5.47E-13 | 6.83E-13 | 2.96E-13 |
| Liver | 9.28E-12 | 6.44E-12 | 6.01E-12 | 4.22E-12 |
| Spleen | 3.54E-13 | 2.35E-13 | 2.52E-13 | 1.62E-13 |
| Bladder (wall) | 8.00E-14 | 2.98E-14 | 1.36E-14 | 2.74E-14 |
| Small intestine (wall) | 4.23E-13 | 3.00E-13 | 2.90E-13 | 2.51E-13 |
| Esophagus | 9.14E-13 | 5.73E-13 | 6.89E-13 | 2.16E-13 |
| Pancreas | 1.59E-12 | 1.07E-12 | 9.45E-13 | 4.61E-13 |
| Thymus | 2.79E-13 | 1.89E-13 | 1.70E-13 | 7.15E-14 |
| Kidneys | 1.21E-12 | 7.71E-13 | 7.16E-13 | 4.20E-13 |
| Large intestine | 3.90E-13 | 2.07E-13 | 3.24E-13 | 4.58E-13 |
| Lungs | 6.69E-13 | 5.29E-13 | 5.13E-13 | 2.46E-13 |
| Eyes | 2.77E-14 | 1.36E-14 | 1.21E-14 | 5.18E-15 |
| Adrenals | 1.72E-12 | 1.08E-12 | 1.24E-12 | 4.62E-13 |
| Remaining organs | 2.39E-13 | 1.22E-13 | 1.12E-13 | 9.31E-14 |

**Supplemental Table 4**. S-value for the organ of the source of the Urinary bladder contents in units of mGy/Bq.

| Organ | 4y | 8y | 11y | 14y |
|---|---|---|---|---|
| Adipose | 2.34E-13 | 1.89E-13 | 1.70E-13 | 9.55E-14 |
| Skin | 1.24E-13 | 7.42E-14 | 6.13E-14 | 4.63E-14 |
| Salivary glands | 2.85E-15 | 8.13E-16 | 3.79E-16 | 3.42E-16 |
| Stomach (wall) | 1.24E-13 | 4.60E-14 | 1.60E-14 | 3.61E-14 |
| Pituitary Gland | 5.82E-16 | 2.47E-17 | 0 | 1.36E-16 |
| Tongue | 3.80E-15 | 9.44E-16 | 3.19E-16 | 3.95E-16 |
| Tonsil | 3.47E-15 | 8.22E-16 | 3.64E-16 | 3.25E-16 |
| Brain | 8.65E-16 | 2.27E-16 | 1.16E-16 | 8.61E-17 |
| colon (wall) | 1.26E-12 | 1.15E-12 | 6.39E-13 | 1.07E-13 |
| ET2 (larynx and pharynx) and Trachea and Bronchi | 6.91E-15 | 1.78E-15 | 9.71E-16 | 1.08E-15 |
| Gall Bladder (wall) | 1.60E-13 | 4.74E-14 | 1.97E-14 | 3.90E-14 |
| active marrow (red marrow) | 2.99E-13 | 2.23E-13 | 2.07E-13 | 1.69E-13 |
| Thyroid | 6.08E-15 | 1.52E-15 | 9.54E-16 | 1.11E-15 |
| Heart | 2.72E-14 | 1.22E-14 | 4.32E-15 | 5.00E-15 |
| Liver | 8.24E-14 | 2.95E-14 | 1.31E-14 | 2.75E-14 |
| Spleen | 6.37E-14 | 2.09E-14 | 1.16E-14 | 9.01E-14 |
| Bladder (wall) | 1.25E-11 | 2.17E-11 | 1.22E-11 | 3.76E-12 |
| Small intestine (wall) | 5.37E-13 | 2.55E-13 | 2.20E-13 | 5.461E-13 |
| Esophagus | 2.24E-14 | 7.65E-15 | 5.47E-15 | 4.76E-15 |
| Pancreas | 1.51E-13 | 4.21E-14 | 2.66E-14 | 5.49E-14 |
| Thymus | 9.86E-15 | 5.10E-15 | 1.36E-15 | 1.50E-15 |
| Kidneys | 1.37E-13 | 5.31E-14 | 2.54E-14 | 3.48E-14 |
| Large intestine | 1.66E-12 | 1.33E-12 | 1.16E-12 | 3.38E-13 |
| Lungs | 2.14E-14 | 7.57E-15 | 3.83E-15 | 5.34E-15 |
| Eyes | 1.77E-15 | 5.50E-16 | 2.12E-16 | 1.154E-16 |
| Adrenals | 8.81E-14 | 3.21E-14 | 1.63E-14 | 1.63E-14 |
| Remaining organs | 3.13E-13 | 2.52E-13 | 1.95E-13 | 1.33E-13 |

**Supplemental Table 5.** S-value for the organ of the source of the Spleen in units of mGy/Bq.s

| Organ | 4y | 8y | 11y | 14y |
|---|---|---|---|---|
| Adipose | 1.34E-13 | 6.64E-14 | 6.32E-14 | 9.24E-14 |
| Skin | 1.55E-13 | 1.01E-13 | 8.76E-14 | 5.49E-14 |
| Salivary glands | 5.20E-14 | 2.88E-14 | 3.08E-14 | 4.09E-15 |
| Stomach (wall) | 2.28E-12 | 1.41E-12 | 1.62E-12 | 1.04E-12 |
| Pituitary Gland | 2.66E-14 | 1.12E-14 | 1.05E-14 | 1.48E-15 |
| Tongue | 6.38E-14 | 3.55E-14 | 3.20E-14 | 4.54E-15 |
| Tonsil | 6.37E-14 | 3.26E-14 | 2.87E-14 | 4.06E-15 |
| Brain | 1.70E-14 | 8.92E-15 | 7.79E-15 | 9.51E-16 |
| colon (wall) | 3.38E-13 | 4.12E-13 | 3.03E-13 | 1.18E-12 |
| ET2 (larynx and pharynx) and Trachea and Bronchi | 1.59E-13 | 8.40E-14 | 8.90E-14 | 1.32E-14 |
| Gall Bladder (wall) | 2.75E-13 | 1.44E-13 | 1.77E-13 | 1.60E-13 |
| active marrow (red marrow) | 2.14E-13 | 1.42E-13 | 1.37E-13 | 7.32E-14 |
| Thyroid | 1.35E-13 | 6.39E-14 | 1.07E-13 | 1.38E-14 |
| Heart | 5.50E-13 | 4.61E-13 | 3.54E-13 | 6.98E-14 |
| Liver | 3.55E-13 | 2.35E-13 | 2.53E-13 | 1.61E-13 |
| Spleen | 7.24E-11 | 5.15E-11 | 4.01E-11 | 3.34E-11 |
| Bladder (wall) | 6.02E-14 | 2.10E-14 | 1.15E-14 | 8.75E-14 |
| Small intestine (wall) | 5.13E-13 | 3.04E-13 | 2.19E-13 | 7.21E-13 |
| Esophagus | 5.05E-13 | 4.23E-13 | 4.67E-13 | 7.08E-14 |
| Pancreas | 9.47E-13 | 1.26E-12 | 1.40E-12 | 2.22E-12 |
| Thymus | 1.96E-13 | 1.82E-13 | 1.52E-13 | 1.83E-14 |
| kidneys | 1.99E-12 | 1.33E-12 | 1.28E-12 | 1.05E-12 |
| Large intestine | 2.95E-13 | 3.41E-13 | 2.49E-13 | 1.02E-12 |
| Lungs | 7.31E-13 | 4.60E-13 | 5.37E-13 | 7.36E-14 |
| Eyes | 2.06E-14 | 1.18E-14 | 1.06E-14 | 1.20E-15 |
| Adrenals | 2.08E-12 | 1.31E-12 | 9.20E-13 | 2.74E-13 |
| Remaining organs | 2.90E-13 | 1.74E-13 | 1.45E-13 | 1.21E-13 |

**Supplemental Table 6.** S-value for the organ of the source of the Total body (excluding urinary bladder contents) in units of mGy/Bq.s

| Organ | 4y | 8y | 11y | 14y |
|---|---|---|---|---|
| Adipose | 3.73E-13 | 2.38E-13 | 2.03E-13 | 1.48E-13 |
| Skin | 1.65E-13 | 1.06E-13 | 9.24E-14 | 7.06E-14 |
| Salivary glands | 4.21E-13 | 2.34E-13 | 2.14E-13 | 1.56E-13 |
| Stomach (wall) | 5.34E-13 | 6.77E-13 | 3.28E-13 | 2.51E-13 |
| Pituitary Gland | 5.02E-13 | 2.68E-13 | 2.49E-13 | 1.71E-13 |
| Tongue | 4.36E-13 | 2.56E-13 | 2.34E-13 | 1.72E-13 |
| Tonsil | 4.72E-13 | 2.73E-13 | 2.48E-13 | 1.82E-13 |
| Brain | 4.32E-13 | 2.43E-13 | 2.25E-13 | 1.58E-13 |
| colon (wall) | 6.25E-13 | 3.93E-13 | 2.93E-13 | 2.40E-13 |
| ET2 (larynx and pharynx) and Trachea and Bronchi | 4.93E-13 | 2.86E-13 | 2.62E-13 | 1.95E-13 |
| Gall Bladder (wall) | 5.21E-13 | 3.33E-13 | 3.12E-13 | 2.37E-13 |
| active marrow (red marrow) | 3.73E-13 | 2.37E-13 | 2.09E-13 | 1.54E-13 |
| Thyroid | 4.96E-13 | 2.60E-13 | 2.80E-13 | 1.98E-13 |
| Heart | 5.39E-13 | 3.40E-13 | 3.09E-13 | 2.34E-13 |
| Liver | 5.20E-13 | 3.36E-13 | 3.03E-13 | 2.24E-13 |
| Spleen | 5.01E-13 | 3.22E-13 | 2.86E-13 | 2.30E-13 |
| Bladder (wall) | 4.71E-13 | 3.53E-13 | 2.91E-13 | 1.93E-13 |
| Small intestine (wall) | 5.25E-13 | 3.48E-13 | 2.93E-13 | 2.21E-13 |
| Esophagus | 5.47E-13 | 3.37E-13 | 3.23E-13 | 2.37E-13 |
| Pancreas | 5.65E-13 | 3.66E-13 | 3.32E-13 | 2.51E-13 |
| Thymus | 4.98E-13 | 3.03E-13 | 2.98E-13 | 2.19E-13 |
| Kidneys | 5.24E-13 | 3.38E-13 | 2.98E-13 | 2.31E-13 |
| Large intestine | 6.14E-13 | 4.21E-13 | 2.95E-13 | 2.43E-13 |
| Lungs | 8.37E-13 | 5.50E-13 | 4.98E-13 | 4.01E-13 |
| Eyes | 3.75E-13 | 2.06E-13 | 1.84E-13 | 1.28E-13 |
| Adrenals | 5.74E-13 | 3.68E-13 | 3.33E-13 | 2.53E-13 |
| Remaining organs | 4.23E-13 | 2.67E-13 | 2.28E-13 | 1.72E-13 |

**Supplemental Table 7.** S-value for the organ of the source of the Total body (excluding urinary bladder contents and kidneys) in units of mGy/Bq.s

| Organ | 4y | 8y | 11y | 14y |
|---|---|---|---|---|
| Adipose | 3.75E-13 | 2.39E-13 | 2.04E-13 | 9.39E-14 |
| Skin | 1.66E-13 | 1.07E-13 | 9.25E-14 | 6.95E-14 |
| Salivary glands | 4.23E-13 | 2.35E-13 | 2.16E-13 | 1.05E-13 |
| Stomach (wall) | 5.33E-13 | 6.77E-13 | 3.26E-13 | 1.84E-13 |
| Pituitary Gland | 5.05E-13 | 2.75E-13 | 2.52E-13 | 1.21E-13 |
| Tongue | 4.39E-13 | 2.58E-13 | 2.35E-13 | 1.23E-13 |
| Tonsil | 4.72E-13 | 2.76E-13 | 2.51E-13 | 1.31E-13 |
| Brain | 4.34E-13 | 2.45E-13 | 2.27E-13 | 1.08E-13 |
| colon (wall) | 6.25E-13 | 3.93E-13 | 2.92E-13 | 1.71E-13 |
| ET2 (larynx and pharynx) and Trachea and Bronchi | 4.94E-13 | 2.87E-13 | 2.63E-13 | 1.46E-13 |
| Gall Bladder (wall) | 5.20E-13 | 3.34E-13 | 3.11E-13 | 1.83E-13 |
| active marrow (red marrow) | 3.74E-13 | 2.37E-13 | 2.10E-13 | 1.54E-13 |
| Thyroid | 4.96E-13 | 2.63E-13 | 2.81E-13 | 1.48E-13 |
| Heart | 5.40E-13 | 3.40E-13 | 3.10E-13 | 1.84E-13 |
| Liver | 5.16E-13 | 3.33E-13 | 3.01E-13 | 1.73E-13 |
| Spleen | 4.93E-13 | 3.17E-13 | 2.80E-13 | 1.76E-13 |
| Bladder (wall) | 4.73E-13 | 3.54E-13 | 2.93E-13 | 2.66E-13 |
| Small intestine (wall) | 5.22E-13 | 3.45E-13 | 2.92E-13 | 1.69E-13 |
| Esophagus | 5.48E-13 | 3.37E-13 | 3.23E-13 | 1.86E-13 |
| Pancreas | 5.60E-13 | 3.59E-13 | 3.24E-13 | 1.94E-13 |
| Thymus | 5.00E-13 | 3.03E-13 | 2.98E-13 | 1.68E-13 |
| Kidneys | 3.11E-13 | 2.09E-13 | 1.85E-13 | 1.52E-13 |
| Large intestine | 6.14E-13 | 4.21E-13 | 2.94E-13 | 1.71E-13 |
| Lungs | 8.41E-13 | 5.52E-13 | 4.99E-13 | 1.90E-13 |
| Eyes | 3.77E-13 | 2.07E-13 | 1.85E-13 | 7.93E-14 |
| Adrenals | 5.59E-13 | 3.50E-13 | 3.20E-13 | 1.93E-13 |
| Remaining organs | 4.23E-13 | 2.68E-13 | 2.29E-13 | 1.22E-13 |